\documentclass{kapproc} 
\upperandlowercase
\newcommand{\lap}{\lower.5ex\hbox{$\; \buildrel < \over \sim \;$}}
\newcommand{\gap}{\lower.5ex\hbox{$\; \buildrel > \over \sim \;$}}

\begin{document}

\articletitle{Concluding Remarks\footnote{In Multi-Wavelength Cosmology, Mykonos, June 2003}}

\author{P. J. E. Peebles}

\affil{Joseph Henry Laboratories\\
Princeton University\\
Princeton NJ 08544 USA}
\email{pjep@princeton.edu}

\chaptitlerunninghead{Concluding Remarks}
\anxx{Peebles, P. J. E.}Ä

\begin{abstract}
The mood at this conference is summarized in David Hughes' comment, ``this decade will be amazing.'' We've just had a pretty good ten years of advances in cosmology and extragalactic astronomy; why should we expect a repeat, another decade of comparable or even greater progress? The obvious answer is that there still are many more questions than answers in cosmology, and a considerable number of the questions will be addressed by research programs planned and in progress: we certainly are going to learn new things. But beyond that is the fact that there is no practical limit to the hierarchy of interesting topics to explore in this subject. I organize my comments on the state of research, and the prospects for substantial new developments in the coming decade of multi-wavelength cosmology, around the concept of social constructions.
\end{abstract}


\section{}

Social scientists inform us that the alpha members of a community set the social standards and constructions, enforce them by the weight of their authority, and see to it that the young members of the community are taught the standards so they will be remembered and enforced by the next generation of alphas. You have experienced all this in your careers in physical science. There is one minor  difference --  we replace the phrase ``social construction'' with ``working hypothesis'' -- and one big addition -- the remarkable fact that some hypotheses become so thoroughly checked as to be convincing approximations to reality. These  comments on the social constructions of cosmology include elements of their history, the present state of the promotions from constructions to established facts, and the prospects for continued additions to our understanding of the real world. 

In their book, {\it The Classical Theory of Fields}, Landau and Lifshitz offer a very sensible caution about the assumption that  the universe is close to homogeneous and isotropic on the scale of the Hubble length. When this book was published, in the 1940s, the evidence for homogeneity was sparse at best: this was a social construction. Now the observational tests are tight and believable. Einstein was led to the picture of homogeneity by his reading of Mach's principle: he felt there had to be matter everywhere to fix inertial motion everywhere. This argument from a philosophical concept led Einstein to an aspect of reality. It is a mystery whether Einstein found the right picture for the right reason. 

Landau and Lifshitz assume without discussion that general relativity theory applies on the scales of cosmology, which is fair enough in a survey of theoretical physics. But at the time -- the first revision of the Russian edition was published in 1948 -- there was just one precision test of the theory, the precession of the perihelion of Mercury, and hints of two others, the gravitational redshift and deflection of light. It certainly made sense to consider the application of the theory to cosmology, but not to trust it. 

The searching probes of gravity physics from the tests that commenced in the 1960s give convincing evidence that general relativity theory is a good approximation on length scales ranging from the laboratory to the size of the Solar System, let us say to $10^{13}$~cm. The Hubble length, $cH_o^{-1}\sim 10^{28}$~cm, is fifteen orders of magnitude larger. This spectacular extrapolation is a social construction, until checked, which is the purpose of the cosmological tests. 

The results certainly look promising. An example is the broad concordance of evidence that the matter density parameter is in the range $0.15\lap\Omega _m\lap0.3$, from analyses of galaxy peculiar velocities, gravitational lensing measurements, the SNeIa redshift-magnitude relation, the cluster baryon mass fraction, the galaxy two-point correlation function, the cluster mass function and evolution, and the ratio $H_ot_o$ of stellar evolution and expansion time scales. A recent addition to the list comes from the wonderfully successful  comparison of the theory and measurements of the anisotropy of the 3K thermal background radiation. This is a demanding test of the gravity theory, which has to describe the propagation of irregularities in the radiation distribution through strongly curved spacetime during the expansion factor $z_{\rm dec}\sim 1000$ since the last substantial interaction of matter and radiation, from initial conditions that have to agree with what grew into the structures observed in the distribution of gas at $z\sim 3$ -- in the Lyman$\alpha$ forest --  and in the present distribution of galaxies. This has given a new check on $\Omega _m$, from the apparent detection of a contribution to the temperature anisotropy from the matter distribution at modest redshifts, an effect demanded by the theory if $\Omega _m$ is significantly below unity. 

Each of these measures of the mean mass density is subject to the hazards of interpretation in astronomy. But it is hard to see how systematic errors could affect the many entries in this list all in the same way. Each measure depends on the assumed physics of gravity and the dark sector, which we are supposed to be testing. The test is the consistency: if we were using the wrong physics the broad concordance would be unlikely. The important thing from the point of view of the cosmological tests is not the value of $\Omega _m$ but rather the convergence of evidence that the estimates of this number are not seriously confused by systematic errors in the observations or by flaws in the underlying theory: we have a good approximation to one aspect of the real world. Physical science can't explain why reality is a meaningful concept, but we can  produce examples of approximations to it. This now includes the measurement of $\Omega _m$.

The evidence that the physics of the standard Friedmann-Lema\^\i tre CDM cosmology is on the right track is a considerable advance, but incomplete. An assignment for this decade is to put the tests of gravity physics applied to cosmology on a systematic basis, in analogy with the program of tests of general relativity on the scale of the Solar System and smaller, though one would of course have to replace the parametrized  framework of that  program with a framework -- maybe parametrized -- that is adapted to what is relevant to cosmology. 

The cosmological principle and general relativity theory are examples of  deep advances in physical science that grew out of concepts of philosophy and elegance, which is why we pay attention to such ideas. The lesson is slippery, of course, because our ideas of what is elegant are adaptable. If the cosmological tests had favored the Steady State cosmological model we would be celebrating the perceptive foresight of a different group of alphas. And recall the history of opinions of Einstein's cosmological constant, $\Lambda$. Einstein came to quite dislike it. Pauli agreed. And Landau and Lifshitz (in the 1951 English translation by M. Hammermesh) asserted that ``it has finally become clear that there is no basis whatsoever'' for the introduction of this term. Others at the time paid no attention to this impressive list of alphas, and they seem to have been on the right track: now there is serious evidence for the detection of $\Lambda$ -- or a term in the stress-energy tensor that acts like it. 

Although most of us would agree that the universe could have done without $\Lambda$, the dark sector of the $\Lambda$CDM cosmology is strikingly simple: the dark energy density is close to constant and the dark matter collects in nearly smooth halos by the gravitational growth of small Gaussian departures from a homogeneous primeval dark matter distribution. This picture for the dark matter was introduced two decades ago, and for some years was one of a  half dozen viable models for galaxy formation. We had useful analytic solutions for explosion models, but serious challenges in an analysis of the physics of a real cosmic explosion. A reliable analysis of the behavior of cosmic strings or monopoles or textures is even more difficult. The CDM model is easy: structure is dominated by particles that move on geodesics, which are readily simulated in numerical computations, and there is the added advantage that structure forms later than in isocurvature variants, so an interesting numerical simulation need not deal with a large expansion factor. Simplicity recommended the CDM model. Now we have substantial observational evidence that it is a useful approximation to what actually happened. 

Is the CDM model complete? One line of thought is that since the dark matter consists of the particles that happen to interact too weakly to be readily observable the dark matter is of course well described as a gas of weakly interacting particles. Another is that the real world seldom is that simple, but that it makes sense to start with the simplest working hypothesis we can get away with, which we will plan to use as a basis for the search for a better approximation, which might in turn lead to a still more complete theory. This is how the physics of the visible sector was discovered.

The search for ideas about how the CDM model might be made more complete -- if that is required -- can be compared to what was happening in the 1930s when Fermi, Yukawa, and others were trying out ideas of how elementary particles interact. Ideas then and now may be represented by Lagrangian densities with forms like 
\[ L = {1\over 2}\phi _{,\nu }\phi ^{,\nu } -V(\phi) +   \sum _a \left[ i\bar\psi _a\gamma\cdot\partial\psi _a
- y_a  (\phi -\phi _a )\bar\psi _a \psi _a \right] . \]
Yukawa's scalar field ($\phi$ in this equation) is complex -- charged -- in order to  couple neutrons and protons. Data were sparse in the 1930s, but Yukawa did know that a reasonable interaction length for nuclear physics -- comparable to the size of an atomic nucleus -- would follow from the potential $V(\phi )= \mu ^2\phi ^2/2$, where the meson mass $\mu$ is about 200 times that of an electron. The standard model for particle physics follows Yukawa, with considerable elaborations. The search for models for the dark sector is at an even simpler level than Yukawa. In the above Lagrangian the scalar field $\phi$ is real, so the Yukawa interaction  $y_a  (\phi -\phi _a )\bar\psi _a \psi _a$  just changes the momenta of dark matter particles (represented by the spin-1/2 field operator $\psi _a$ for the $a^{\rm th }$ family, where $y_a$ and $\phi _a$ are real constants). If the  potential $V(\phi )$ in the dark sector is close to Yukawa's form, and $\mu$ is relatively large, this is a model for the self-interacting cold dark matter picture. If $V$ varies only  slowly with $\phi$ the $a^{\rm th}$ family behaves as particles with variable mass, $m_a=y_a (\phi -\phi _a)$, and the gradient of the mass is a fifth force -- a long-range inverse square force of attraction of dark matter particles that adds to the gravitational attraction. This physics traces back to Nordstr\"om's (1912) scalar field model for gravity in Minkowski spacetime, from there to the scalar-tensor gravity theories that were much discussed in the 1950s and 1960s, and from there to the present-day ideas of  dilaton fields that would have observable effects -- variable parameters -- in the visible sector, and maybe a considerably stronger fifth force in the dark sector. A potential energy density $V(\phi )$ that varies slowly with $\phi$ also appears in a popular model for the dark energy or quintessence. The pedigree is impressive, and it suggests many scenarios for physics in the dark sector even without elaborations comparable to what happened to the model for particle physics after Fermi and Yukawa. To be seen is whether it might lead us to a  model that can remedy apparent anomalies -- some of which are mentioned below --  in the standard $\Lambda$CDM cosmology.

If the present standard cosmology really differs from reality enough to matter it will appear in anomalies. But there is a problem, that we cannot in practice unambiguously connect given physics and initial conditions to the details of cosmic structure that are revealed in the observations. How do we decide whether apparent anomalies are only the result of the difficulty of modeling the physics, or whether real failures of the theory have been obscured by the modeling? We ned a new generation of tests of reliability of the hypotheses that are used to model the connection between the theory and observations.  The situation is similar to condensed matter physics, where complexity also drives model building, but very different in the sense that we have excellent reason to trust the underlying physics of condensed matter. We will gain confidence in the physics of the dark sector by the accumulation of tests, including the examination of alternatives to CDM. This is another assignment for multi-wavelength cosmology.

Two apparent anomalies that fascinate me have to do with the cosmic web and the galaxy merger rate. The cosmic web is a striking visual feature of numerical simulations of the CDM model, and the web does predict the observed walls of galaxies. But in maps of halo distributions in the simulations I see chains of dwarfs running into the voids between the concentrations of the more massive halos, which I don't see in maps of the real galaxy distribution. Maybe this is a result of the complexity of modeling the connection between theory and observations, exploration of which is part of the research assignment. For now I'm counting the cosmic web of galaxies as a social construction. 

I hasten to emphasize that I am deeply impressed by the successful account the cosmic web of gas offers for the  statistical measures of the Lyman$\alpha$ forest. On the other hand, I wish I felt better about the  apparent lack of disturbance by whatever added heavy elements to the hydrogen in the Lyman$\alpha$ forest clouds.

Another apparent anomaly is the rate of merging of closely bound galaxies. A pair of galaxies separated by a few half-light radii is routinely labeled a merging system, whether at high redshift or low, whether in a group or a rich cluster of galaxies. There is a good reason -- simulations and analytic estimates predict the pair will merge in a few crossing times -- but is it more than a social construction? The theoretical argument is sound, but only if we have the right physical model for the dark matter, as a nearly collisionless gas, which is not yet something we know. On the observational side, it is often said that the merger remnant of a compact group of spirals is an elliptical, but I also hear that the pattern of element abundances in the progenitors -- typically late types -- does not look like a promising match to the abundances in a typical early-type galaxy. Also puzzling is the effect of mergers on the shape of the low order galaxy correlation functions. The two-point function is a good approximation to a power law from 10~Mpc separation down to separations of a few half-light radii.  Standard estimates of the cosmic merger rate assume close pairs merge in a few orbit times. If so, what preserves the power law form? 

Again, I have to qualify these remarks. There is good evidence of mergers at low redshift: we see a clear example in the Centaurus group, where the big elliptical clearly has recently merged with a dusty galaxy, and there are several other classical examples of galaxies that surely are observed in the act of merging. But these spectacular systems do not seem to be all that common: the familiar examples are repeatedly cited. The assignment is to show whether the number of merging galaxies at low redshift really is consistent with the theoretical prediction that galaxies closer than a few half-light radii merge on time scales small compared to the Hubble time. 

We might consider also that we have to live with quite a few coincidences within the standard $\Lambda$CDM Friedmann-Lema\^\i tre cosmology. Heavily advertised nowadays is the coincidence that we flourish just as the universe is making the transition from matter-dominated expansion to an approximation to the de Sitter solution. Maybe related to this is the observation that we flourish just as the Milky Way is running out of gas for the formation of new generations of planetary systems like our own, along with the rapid collapse of the global star formation rate density since redshift $z=1$. Less widely discussed these days is the possible relation to Dirac's large numbers coincidence: the ratio of the present Hubble length to the classical electron radius is close to another enormous number, the ratio of the electric and gravitational forces between a proton and electron. Another timing coincidence I suppose is unrelated, but also curious, is that in the standard cosmology optically selected galaxies have just now become good mass tracers: the theory seems to predict strong biasing at redshift $z=1$. In the standard cosmology the mass of a large galaxy is dominated by dark matter in the outer parts, and by stars near the center. The conspiracy is that the distributions of these two components produce a net mass density run that shows no feature at the transition from high to low mass-to-light ratio. And finally, the $\Lambda$CDM cosmology predicts separation-dependent bias of light as a tracer of mass: the ratio of the mass autocorrelation function to the two-point correlation function of optically selected galaxies is a function of separation. But it is curious that the galaxies seem to give the better approximations to power law forms for the low order correlation functions deep in the nonlinear clustering limit, rather than the mass that is supposed to control the dynamics. 

It is reasonable to expect that some of these curiosities are nothing more than accidents, and some will be seen not to be curious at all when we have a really good understanding of the theory and its relation to the observations. But it is sensible to be aware of the possibility that some are clues to improvements in the physics. 

What might come from continued multi-wavelength research on such challenges to cosmology? I expect the paradigms will continue to rest on the relativistic Friedmann-Lema\^\i tre model, or some good approximation to it, because general relativity theory has passed quite demanding checks on the scales of cosmology. But we owe it to our subject to turn these scattered checks into a systematic survey of the constraints on the physics of spacetime and gravity.

I do not expect a paradigm shift back to the Einstein-de Sitter model: the lines of evidence for low $\Omega _m$ are impressively well checked by many independent applications of the theory that depend on quite different elements of the astronomy. The evidence of detection of the cosmological constant is serious, too, but not as well checked as $\Omega _m$. The $\Lambda$ term has been debated  for more than eight decades; we can wait a few more years before deciding whether it deserves a place in  the list of convincingly established results. 

In the next ten years multi-wavelength observations, including (in the words of a participant) ``millimeter, submillimeter and FIR observations with the
imaging fidelity currently enjoyed by X-ray, optical, IR and radio
astronomers" will produce an enormous increase in our knowledge of cosmic structure, and that is going to drive the development  of esceedingly detailed models to relate the theory to the observations. The theory of choice will continue to be $\Lambda$CDM, unless or until the observations drive us to something better. While waiting to see whether that happens a assignment for model builders is to develop a convincing case for how far they have gone beyond curve fitting.

After a major advance in a physical science, such as we have seen in the past decade in cosmology, there is the tendency to ask whether the subject has now reached completion, requiring only the ``addition of decimal places.'' You don't hear this talk among astronomers, and I wouldn't expect it to be on astronomers' minds in the coming decades, because there is no practical limit to the layers of detail one may study about things like the populations of stars, planetary systems, and civilizations that are communicating by radio broadcasts in the Milky Way, in the Magellanic clouds, and on out. We have good reason to expect the first decade of the 21$^{\rm st}$ century will be remembered as a golden time for cosmology, but we can be sure there will be room for productive applications of multi-wavelength cosmology for decades to come.

My overall conclusion is that you should pay attention to the alphas -- their concepts of simplicity and elegance really have led to deep advances in our understanding of the material world -- but then go make the measurements -- the alphas have feet of clay like everyone else. 

\vspace {7mm}

I am grateful to Manolis Plionis for inspiration, David Hogg and David Hughes for advice, and the USA National Science Foundation for financial support for this essay.

\end{document}